\documentclass[preprint,
preprintnumbers,amsmath,amssymb]{revtex4}
\usepackage{dcolumn}
\usepackage{bm}
\usepackage{graphicx}

\begin{document}

\title{Zero Temperature Holographic Superfluids with Two Competing Orders}

 \author{Ran Li$^1$}
 \thanks{liran@htu.edu.cn}

 \author{Yu Tian$^{2,3}$}
 \thanks{ytian@ucas.ac.cn}

 \author{Hongbao Zhang$^{4,5}$}
 \thanks{hzhang@vub.ac.be}

 \author{Junkun Zhao$^1$}
 \thanks{zhaojkun1991@163.com}

\affiliation{
 $^1$ Department of Physics, Henan Normal University, Xinxiang 453007, China\\
 $^2$ School of Physics, University of Chinese Academy of Sciences, Beijing 100049, China\\
 $^3$ Shanghai Key Laboratory of High Temperature Superconductors, Shanghai 200444, China\\
 $^4$ Department of Physics, Beijing Normal University, Beijing 100875, China\\
 $^5$ Theoretische Natuurkunde, Vrije Universiteit Brussel, and \\
      The International Solvay Institutes, Pleinlaan 2, B-1050 Brussels, Belgium}

\begin{abstract}

We initiate the investigation of the zero temperature holographic superfluids with two competing orders, where besides the vacuum phase, two one component superfluid phases, the coexistent superfluid phase has also been found in the AdS soliton background for the first time. We construct the complete phase diagram in the $e-\mu$ plane by numerics, which is consistent with our qualitative analysis. Furthermore, we calculate the corresponding optical conductivity and sound speed by the linear response theory. The onset of pole of optical conductivity at $\omega=0$ indicates that the spontaneous breaking phase always represents the superfluid phase, and the residue of pole is increased with the chemical potential, which is consistent with the fact that the particle density is essentially the superfluid density for zero temperature superfluids. In addition, the resulting sound speed demonstrates the non-smoothness at the critical points as the order parameter of condensate, which indicates that the phase transitions can also be identified by the behavior of sound speed. Moreover, as expected from the boundary conformal field theory, the sound speed saturates to $\frac{1}{\sqrt{2}}$ at the large chemical potential limit for our two component holographic superfluid model.
\end{abstract}

\maketitle

\section{Introduction}

The AdS/CFT correspondence has provided us with a new approach to investigate strongly interacting systems by studying their weakly coupled gravity duals with one extra dimension\cite{M,GKP,W}. In particular, it has recently been used to model various phenomena in condensed matter systems such as superfluids and superconductors\cite{G,HHH1,HHH2}. By putting a charged scalar field on top of the bulk Einstein-Maxwell theory, the AdS black hole becomes unstable to form a scalar hair near a critical temperature, which have a dual interpretation on the boundary as a second order phase transition from a normal fluid to an s-wave superfluid or from a normal metal to an s-wave superconductor. Such a holographic system exhibits many characteristic properties shared by real materials and the basic setup has been extended to describe the condensate with a more sophisticated structure on the boundary by taking into account other types of matter fields in the bulk\cite{GP,DG,CLL,CKMWY,BHRY}.

On the other hand, the insulator to superconductor phase transition at zero temperature has also been implemented by holography in \cite{NRT}, where the AdS soliton will carry a scalar hair when one cranks up the chemical potential to a certain critical value. It is found that this holographic model has a very similarity with the resonating valence bond approach in modeling the high-T$_c$ superconductors. When this phase transition is viewed as a vacuum to superfluid phase transition, the corresponding phase diagram bears a strong resemblance to the one observed recently in ultracold cesium atoms, where the compactified dimension in the AdS soliton background can
be naturally identified as the reduced dimension in optical lattices by the very steep harmonic
potential as both mechanisms make the effective dimension of the system in consideration
reduced in the low energy regime\cite{ZHTC}. By taking into account the backreaction of matter fields onto the metric, the complete phase diagram in $T-\mu$ plane has further been constructed in \cite{HW} by numerics. Near the critical points, the analytical calculations are available and consistent with the numerical results\cite{CLZ}.

The strongly correlated systems also demonstrate the phases with many coexistent orders. Therefore, it is desirable to investigate the various order parameters coexistent in the holographic setup. Actually, such a holographic construction of the competing order parameters has been successfully accomplished in the AdS black hole background in \cite{BHMRS} for the first time. Furthermore, the case for the two competing scalar order parameters has been extensively explored by going beyond the probe limit in \cite{CLLW}. More works on the holographic multi-band systems can be found in \cite{HLM,DGSW,KKS,NCGZ,LCLW,N,CB,DM,AAJML,WYW,MRM,NCGLZ,NZ,MSW}. For a more comprehensive review, please check \cite{CLLY}.

 But nevertheless, it is always significant to see how these orders compete with one another at zero temperature, as it is believed that the quantum criticality can serve as the best point of departure for understanding what happens at finite temperature. However, to the best of our knowledge, the holographic investigation of coexistent phases on top of the AdS soliton geometry has not been touched upon yet. Due to the essential difference between the AdS black hole and AdS soliton, it is a priori unknown whether there are such coexistent phases on top of the AdS soliton geometry. Not only is this issue of academic interest by itself, but also may be relevant to the feasibility of implementation of a mixture of Bose and Fermi superfluids in realistic systems such as liquid helium and cold atoms\cite{Helium1,Helium2,Cold1,Cold2}.

The purpose of this paper is to fill such a gap by investigating the zero temperature holographic superfluid model with the two competing scalar orders in the aforementioned AdS soliton background\cite{footnote}. The paper is structured as follows. In Section \ref{setup}, we will present our zero temperature holographic superfluid model, where the two charged scalar fields are coupled to one single $U(1)$ gauge field on top of the AdS soliton background. Then we shall make some analytical arguments in Section \ref{qualitative} about the possible region for the existence of bulk solution dual to the boundary coexistent phase by rephrasing the problem in terms of a Schrodinger like one. In Section \ref{numerical}, we apply the pseudo-spectral method to numerically solve the the non-linear equations of motion and construct the complete phase diagram by the free energy analysis, where as expected the coexistent phase shows up. After that, in Section \ref{linearresponse} we affirm that the system with the non-vanishing condensates indeed describes the superfluid phase by calculating the optical conductivity in the context of linear response theory. Furthermore, by employing the frequency domain analysis in Section \ref{nmss}, we spot the hydrodynamic normal modes, extract the sound speed from the dispersion relation, and figure out the variation of sound speed with respect to the chemical potential. We conclude our paper with some discussions in Section \ref{end}.

\section{Holographic Model}\label{setup}
In this section, we begin with the holographic setup for the zero temperature two component superfluids, where two complex scalar fields are coupled to one single $U(1)$ Abelian gauge field in the AdS soliton background. The action for the model reads
\begin{equation} \label{action}
S=\frac{1}{16\pi G}\int d^{d+1}x\sqrt{-g}[R+\frac{d(d-1)}{l^2}+L_{matter}].
\end{equation}
Notice that $G$ is the Newton's gravitational constant, the AdS curvature radius $l$ is related to the negative cosmological constant as $\Lambda=-\frac{d(d-1)}{2l^2}$, and Lagrangian of the matter fields is given by $L_{matter}=\frac{l^2}{e_2^2}L$  with $L$ in the form of
\begin{equation}
L=-\frac{1}{4}F^{ab}F_{ab}-|D_1\Psi_1|^2-m_1^2|\Psi_1|^2-|D_2\Psi_2|^2-m_2^2|\Psi_2|^2.
\end{equation}
Here, we are writing $F=dA$, $D_1=\nabla-i\frac{e_1}{e_2}A$, $D_2=\nabla-iA$, with $e_i$ and $m_i$ $(i=1,2)$ the charge and mass carried by the complex scalar field $\Psi_i$, respectively. In what follows, we shall work with the probe limit. Namely, the backreaction of matter fields to the bulk geometry is ignored, which can be achieved by taking the large $e_{2}$ limit but keeping the ratio of two scalar field charges, namely $e=e_1/e_2$, finite.  With this in mind, we can perform the interplay of the matter fields above a fixed background geometry, which is a solution for the vacuum Einstein gravity with the negative cosmological constant. For our purpose, we choose the AdS soliton solution as our background geometry, which is obtained by the double Wick rotation of the planar AdS Schwarzschild black hole as
\begin{equation}
ds^2=\frac{l^2}{z^2}[-dt^2+d\mathbf{x}^2+\frac{dz^2}{f(z)}+f(z)d\chi^2].
\end{equation}
Notice that in this coordinate system $f(z)=1-(\frac{z}{z_0})^d$ with $z=z_0$ the tip where our geometry caps off and $z=0$ the AdS boundary. In order to avoid a conical singularity at the tip, we must impose a periodicity $\frac{4\pi z_0}{d}$ along the $\chi$ coordinate direction. The inverse of the periodicity is interpreted as the confining scale for the dual boundary theory. Without loss of generality, below we take $16\pi G e_2^2=1$, $l=1$, and $z_0=1$. In addition, we shall focus solely on the action of matter fields because the leading $e_2^0$ contribution has been frozen due to the fixed background geometry.

The equations of motion for the matter fields read
\begin{eqnarray}
&&D_{1a}D_1^a\Psi_1-m_1^2\Psi_1=0,\\
&&D_{2a}D_2^a\Psi_2-m_2^2\Psi_2=0,\\
&&\nabla_aF^{ab}=ie\left(\overline{\Psi_1}D_1^b\Psi_1-\Psi_1\overline{D_1^b\Psi_1}\right)+
i\left(\overline{\Psi_2}D_2^b\Psi_2-\Psi_2\overline{D_2^b\Psi_2}\right).
\end{eqnarray}
whence the asymptotical behavior for the matter fields near the AdS boundary is of the form
\begin{eqnarray}\label{conformaldimension}
&&\Psi_1\rightarrow\psi_{-,1}z^{\Delta_{-,1}}+\psi_{+,1}z^{\Delta_{+,1}}\label{asympt1} ,\\
&&\Psi_2\rightarrow\psi_{-,2}z^{\Delta_{-,1}}+\psi_{+,2}z^{\Delta_{+,1}}\label{asympt2} ,\\
&&A_\mu\rightarrow a_\mu+b_\mu z^{d-2}\label{asympt3}.
\end{eqnarray}
with the axial gauge $A_z=0$ and conformal dimension $\Delta_{\pm,i}=\frac{d}{2}\pm\sqrt{\frac{d^2}{4}+m_i^2}$. Below we shall devote ourselves to the case of $m_1^2=0$, $m_2^2=-2$ and $d=3$. Correspondingly, we obtain $\Delta_{-,1}=0,\Delta_{+,1}=3$ and $\Delta_{-,2}=1,\Delta_{+,2}=2$, respectively. According to the holographic dictionary for the standard quantization in the AdS/CFT correspondence, we have
\begin{eqnarray}
\langle j^\mu\rangle&=&\frac{\delta S_{ren}}{\delta a_\mu}=b^\mu,\nonumber\\
\langle O_{+,1}\rangle&=&\frac{\delta S_{ren}}{\delta\psi_{-,1}}=\overline{\psi_{+,1}},\nonumber\\
\langle O_{+,2}\rangle&=&\frac{\delta S_{ren}}{\delta\psi_{-,2}}=\overline{\psi_{+,2}}.\label{dict}
\end{eqnarray}
Here $j^\mu$ is interpreted as the boundary conserved particle current, the expectation value of the scalar operator $O_{+,1}$ and $O_{+,2}$ is regarded as the order parameters of condensate in holographic superfluid system, and $S_{ren}$ represents the holographic renormalized on-shell action obtained by adding the counter terms to remove the divergence of the original action, i.e.,
\begin{eqnarray}
S_{ren}=S+\int d^3x\sqrt{-h}|\partial\Psi_1|^2-\int d^3 x\sqrt{-h}|\Psi_2|^2.
\end{eqnarray}

When either of the two bulk scalar fields takes a non-vanishing profile under the condition that both of the sources $\psi_{-,1}$ and $\psi_{-,2}$ are switched off, the corresponding scalar operator will acquire a non-vanishing expectation value, which corresponds to the U(1) symmetry spontaneous breaking. Then the dual boundary system is perceived to be in a superfluid phase. In particular, when both of the two scalar operators have a non-vanishing expectation value simultaneously, the dual boundary system is in the coexistent superfluid phase.

To make our life easier, in what follows we shall not touch upon the striped phase structure as in \cite{JE,KWG}, but restrict ourselves solely on the homogeneous phase structure of our model, which can be implemented by the homogeneous ansatz for the bulk matter fields as
\begin{equation}
\Psi_1=\Psi_1(z),\quad \Psi_2=\Psi_2(z),\quad A_\mu dx^\mu=A_t(z)dt.
\end{equation}
As a result, the independent equations of motion can be reduced to
\begin{eqnarray}
&&\Psi_1^{''}+\left(\frac{f^{'}}{f}-\frac{2}{z}\right)\Psi_{1}^{'}+
\left(-\frac{m_1^2}{z^2f}+\frac{e^2A_t^2}{f}\right)\Psi_1=0,\label{static1}\\
&&\Psi_2^{''}+\left(\frac{f^{'}}{f}-\frac{2}{z}\right)\Psi_2^{'}+
\left(-\frac{m_2^2}{z^2f}+\frac{A_t^2}{f}\right)\Psi_2=0,\label{static2}\\
&&A_t^{''}+\frac{f^{'}}{f}A_t^{'}-2e^2\frac{\Psi_1^2}{z^2f}A_t-2\frac{\Psi_2^2}{z^2f}A_t=0,
\end{eqnarray}
where the prime denotes the derivative with respect to $z$.

\section{Qualitative Analysis}\label{qualitative}

Before solving our holographic model numerically, we would like to carry out a brief qualitative discussion on the possible solutions to the above equations. First, there obviously exists a trivial solution with a constant gauge potential $A_t=\mu$ and a vanishing profile for both scalar fields, which corresponds to the vacuum phase in the dual boundary field system. On the other hand, when one of the bulk scalars is set to zero, the system is reduced to the one component superfluid with the critical chemical potential $\mu_i$, as studied in \cite{NRT}. But in order to see whether there is a bulk solution dual to a coexistent superfluid phase, we like to convert our equations for scalar fields
\begin{eqnarray}
&&z^4\partial_z\left(\frac{f}{z^2}\partial_z\Psi_1\right)+\left(-m_1^2+e^2z^2A_t^2\right)\Psi_1=0,\\
&&z^4\partial_z\left(\frac{f}{z^2}\partial_z\Psi_2\right)+\left(-m_2^2+z^2A_t^2\right)\Psi_2=0.
\end{eqnarray}
into Schrodinger like ones as
\begin{eqnarray}
&&-\partial_y^2\widetilde{\Psi_1}+V_1\widetilde{\Psi_1}=0,\\
&&-\partial_y^2\widetilde{\Psi_2}+V_2\widetilde{\Psi_2}=0.
\end{eqnarray}
Here we have defined a new variable $y\in[0,y^*]$ as $dy=\frac{1}{\sqrt{f}}dz$ with $y^*=\frac{\sqrt{\pi}\Gamma(\frac{4}{3})}{\Gamma(\frac{5}{6})}\approx 1.40218$ and introduced the new functions $\Psi_1=Y\widetilde{\Psi_1},\Psi_2=Y\widetilde{\Psi_2}$ with $Y=\frac{z}{f^\frac{1}{4}}$. In addition, the potentials are given by
\begin{eqnarray}
&&V_1=-e^2A_t^2+\frac{m_1^2}{z^2}-\frac{z^2\partial_z(\frac{f\partial_zY}{z^2})}{Y},\\
&&V_2=-A_t^2+\frac{m_2^2}{z^2}-\frac{z^2\partial_z(\frac{f\partial_zY}{z^2})}{Y}.
\end{eqnarray}
which are found to have the following asymptotic behaviors
\begin{eqnarray}
&&V_i\approx \frac{m_i^2+2}{y^2}, y\rightarrow 0,\label{uv}\\
&&V_i\approx -\frac{1}{4(y-y^*)^2},y\rightarrow y^*.\label{ir}
\end{eqnarray}
The Hamiltonian for the potential $V(y)=\frac{\kappa}{
y^2}$ is called Calogero Hamiltonian, which leads to an ill-defined Schrodinger problem for $\kappa<-\frac{1}{4}$ due to Landau fall effect, namely the system will be unstable to infinitely many negative energy states. When
applied to (\ref{uv}), this gives rise to the well known BF bound $m^2\geq-\frac{9}{4}$. On the other hand, according to (\ref{ir}), the behavior
near the tip indicates that the system is marginally stable. However, the introduction of the gauge potential lowers the
ground state energy. In particular, when the gauge potential is cranked up to a certain value, the system will become unstable with the
scalar condensed right from the tip all the way to the AdS boundary. This IR instability corresponds to the spontaneous breaking mechanism for the aforementioned one component superfluid phase.

On the basis of the lemma proved in \cite{BHMRS}, we know that if two potentials $V_1$ and $V_2$ are over the same domain with $V_1>V_2$, then the lowest eigenvalue for the Hamiltonian associated with $V_1$ will be strictly greater than the lowest eigenvalue for that associated with $V_2$. This indicates that if the lowest eigenvalue mode for the case of $V_2$ is a zero mode, then $V_1$ cannot give rise to a zero mode in the same region. Back to our holographic model, we have
\begin{equation}
V_1-V_2=\left(1-e^2\right)A_t^2+\frac{2(m_1^2-m_2^2)}{z^2}
\end{equation}
for the two scalar fields. Note that we are focusing on the case of $m_1^2=0$ and $m_2^2=-2$. Thus for $e\leq1$, we always have $V_1>V_2$ and the zero mode of $\Psi_2$ condensates before $\Psi_1$, which further prevents the condensation of $\Psi_1$ by depleting the gauge potential. Hence, the corresponding phase structure is the same as that for the one component holographic superfluid only with $\Psi_2$ condensation. The interesting situation happens to the $e>1$ case, because there is sort of competition between the gauge potential term and mass dependent term. It is reasonable to expect that the coexistent phase shows up in an intermediate region of chemical potential, which will be confirmed by our numerical calculation below. Furthermore,
one may naively think that as we increase the chemical potential the scalar field with a large charge will eventually dominate. However, as demonstrated by our numerics, due to the $\frac{1}{z^2}$ divergent behavior near the AdS boundary, the mass dependent term becomes more important such that the scalar field with a small charge will dominate at the large chemical potential.

\section{Phase Diagram}\label{numerical}

\begin{figure}
\begin{center}
\includegraphics[width=7.5cm]{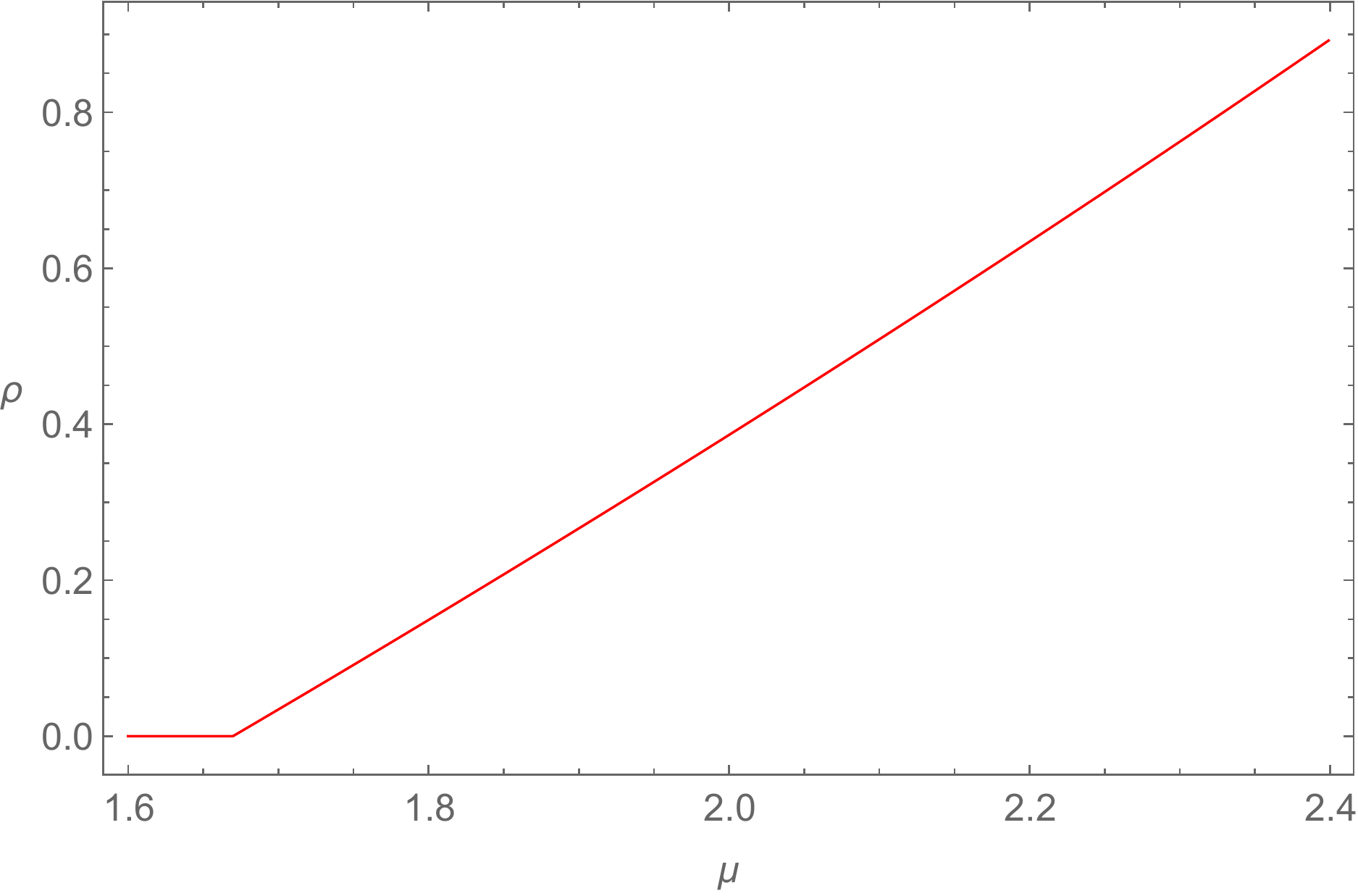}
\includegraphics[width=7.5cm]{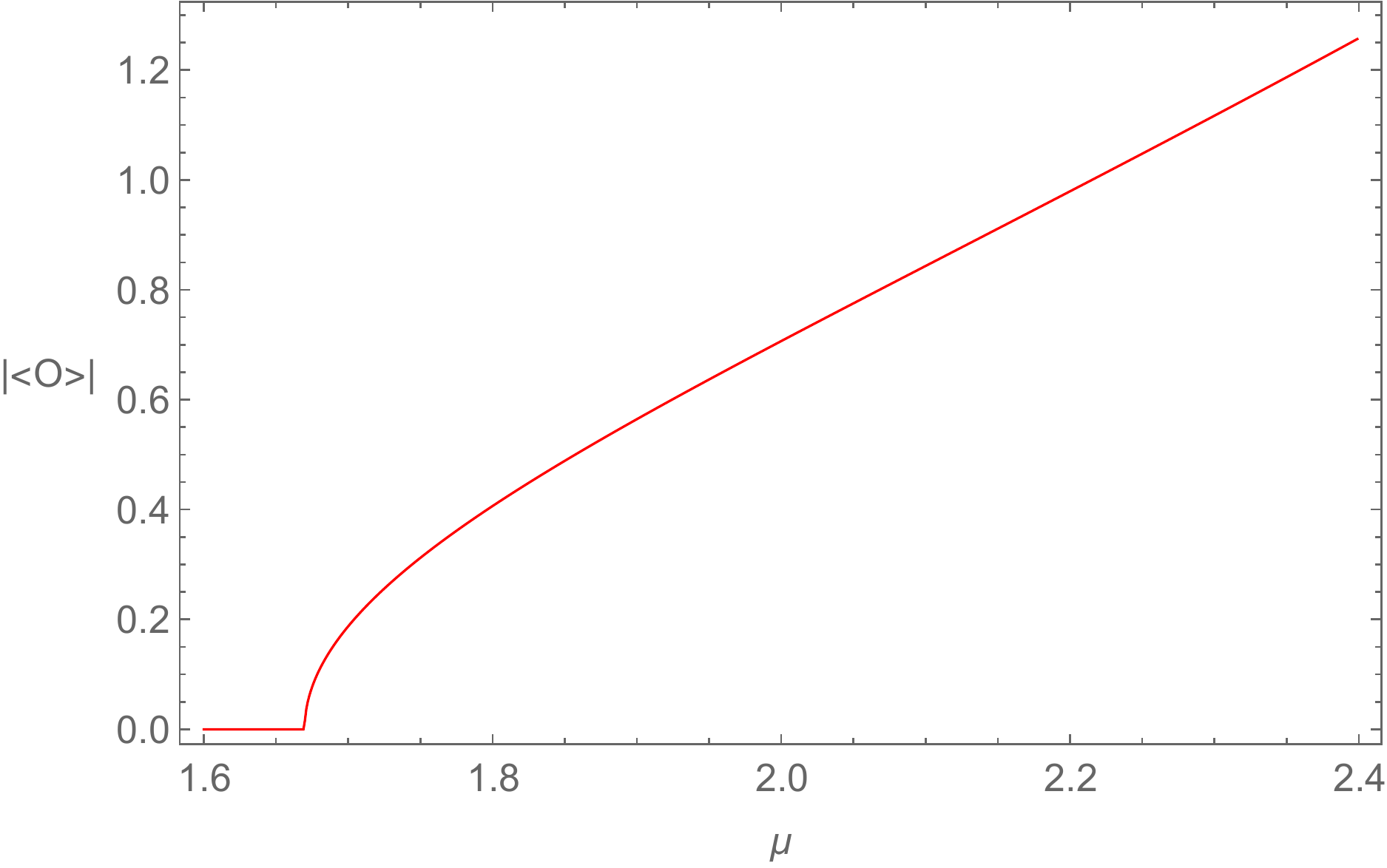}
\includegraphics[width=7.5cm]{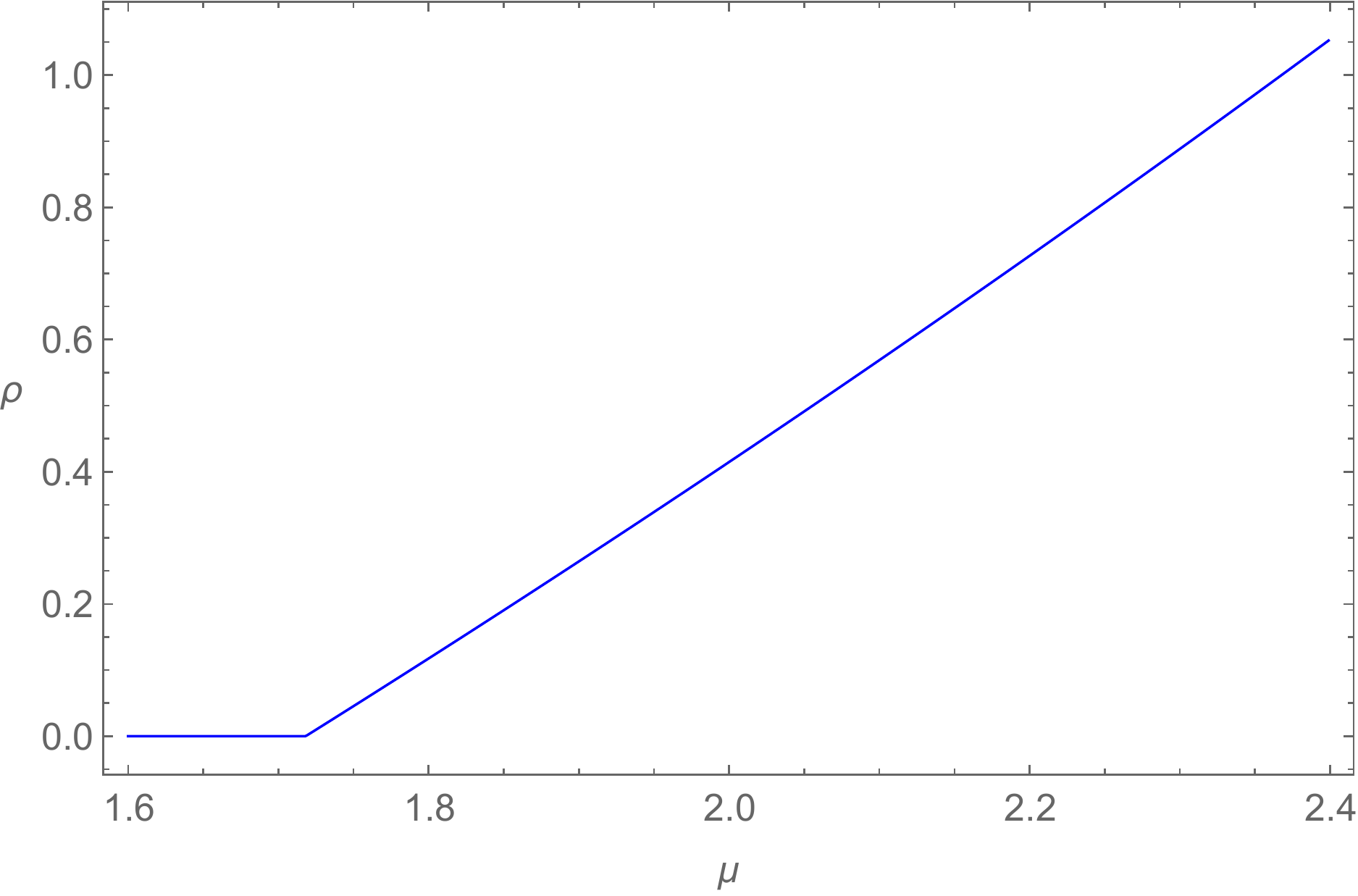}
\includegraphics[width=7.5cm]{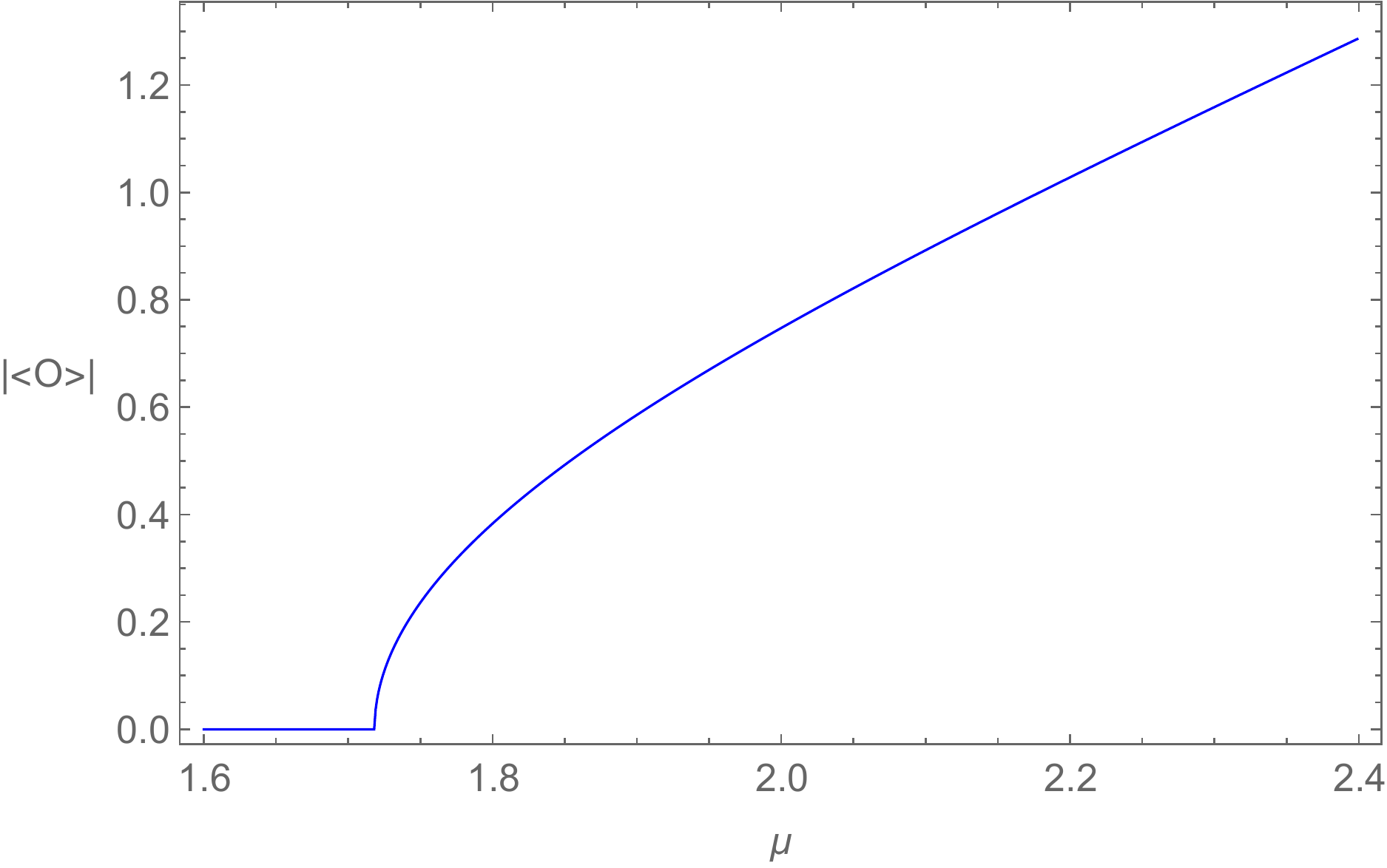}
\end{center}
\caption{The variation of particle density and condensate with respect to the chemical potential $\mu$ in one component holographic superfluid, where the top is for $\Psi_1$ with the second order phase transition triggered at $\mu_1=1.669$, and the bottom is for $\Psi_2$ with the second phase transition occurred at $\mu_2=1.718$.}
\label{phase1}
\end{figure}

\begin{figure}
\begin{center}
\includegraphics[width=7.5cm]{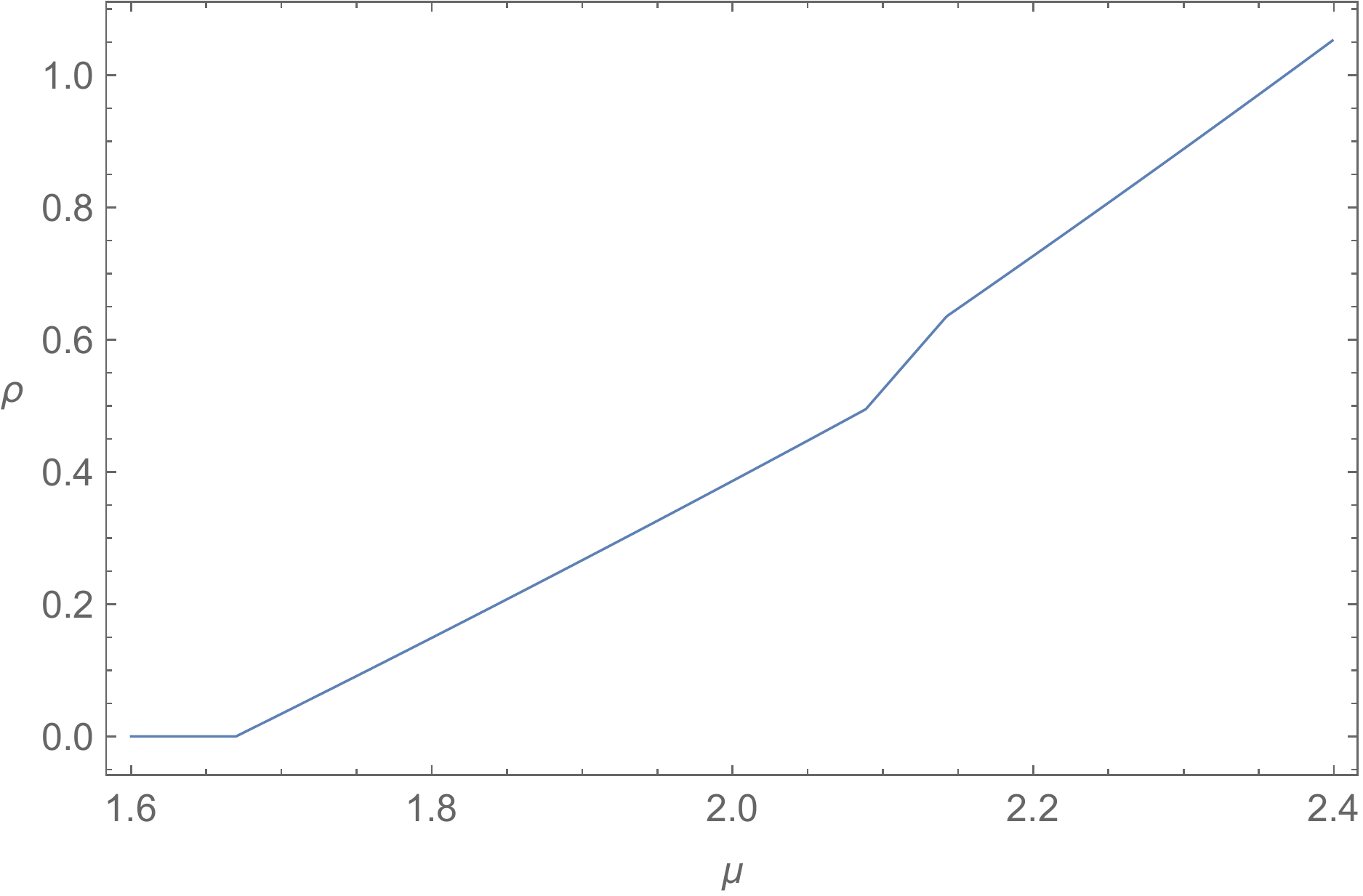}
\includegraphics[width=7.5cm]{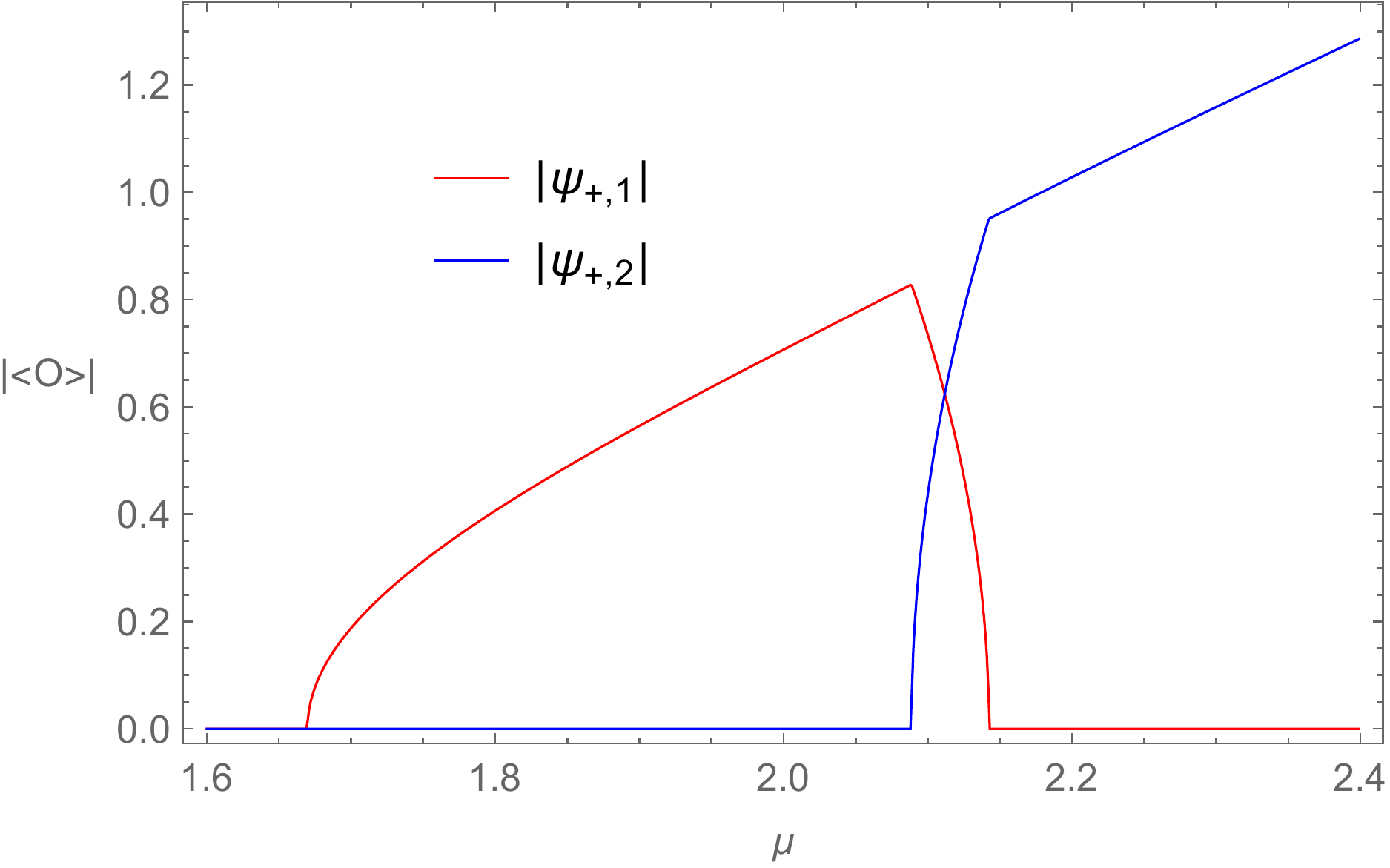}
\end{center}
\caption{The variation of particle density and condensate with respect to the chemical potential in two component holographic superfluid, where the chemical potential for the coexistent superfluid phase with two competing orders lies in the region between $\mu_{c1}=2.088$ and $\mu_{c2}=2.143$.}
\label{phase2}
\end{figure}

In this section, we shall figure out the phase diagram for our holographic model by numerics. As alluded to in the previous section, the rich phenomenon happens to the case of $e>1$. So we would like to take $e=1.63$ as a concrete example for the purpose of demonstration. To this end, we first rewrite the bulk fields in the following form
\begin{equation}
\Psi_1=\psi_1(z),\quad \Psi_2=z\psi_2(z).
\end{equation}
As a result, the equations of motion for the static configurations can be expressed as
\begin{eqnarray}
&&z(1-z^3)\psi_1^{''}-(2+z^3)\psi_1^{'}+ze^2A_t^2\psi_1=0,\\
&& (1-z^3)\psi_2^{''}-3z^2\psi_2^{'}+(A_t^2-z)\psi_2=0,\\
&&z^2(1-z^3)A_t^{''}-3z^4A_t^{'}-2(e^2\psi_1^2+z^2\psi_2^2)A_t=0.
\end{eqnarray}
These coupled non-linear differential equations together with Dirichlet boundary conditions at the AdS boundary
 \begin{equation}
 \psi_1=\psi_2=0, A_t=\mu
 \end{equation}
 can be solved by the pseudo-spectral method, supplemented with Newton-Raphson iteration method. Then we can read off all the physical quantities of interest from the resulting bulk matter field configurations by the holographic dictionary (\ref{dict}).

We plot below the variation of particle density and condensate with respect to the chemical potential for one component holographic superfluid in Figure \ref{phase1}. As we see from both the behaviors of particle density and condensate, the system undertakes a second order phase transition from the vacuum to superfluid phase. In addition, the critical chemical potential for $\Psi_1$ is less than that for $\Psi_2$, which turns out to be both the sufficient and necessary conditions for the emergence of coexistent superfluid phase. As shown in Figure \ref{phase2} for our two component holographic superfluid, when one cranks up the chemical potential to a certain critical chemical potential $\mu_{c1}$, the coexistent phase shows up, where the condensate of $\Psi_1$ starts to decrease, accompanied by the emergence of the condensate of $\Psi_2$. Eventually the competition between these two orders ends at another critical chemical potential $\mu_{c2}$, where the condensate of $\Psi_1$ disappears with the only occurrence of the condensate of $\Psi_2$. As evident from Figure \ref{phase2}, the phase transitions occuring at $\mu_{c1}$ and $\mu_{c2}$ are both second order.

\begin{figure}
\begin{center}
\includegraphics[width=7.5cm]{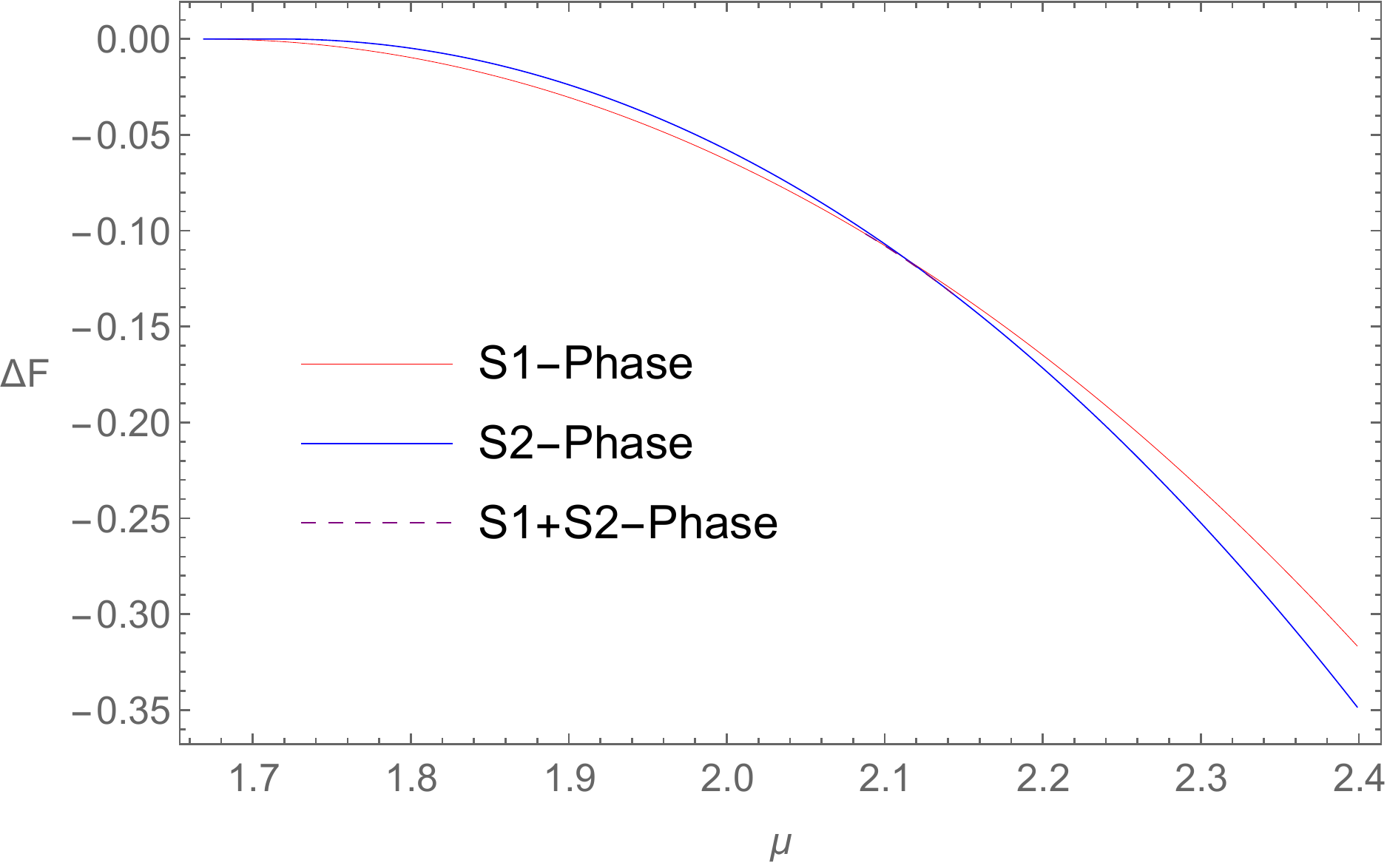}
\includegraphics[width=7.5cm]{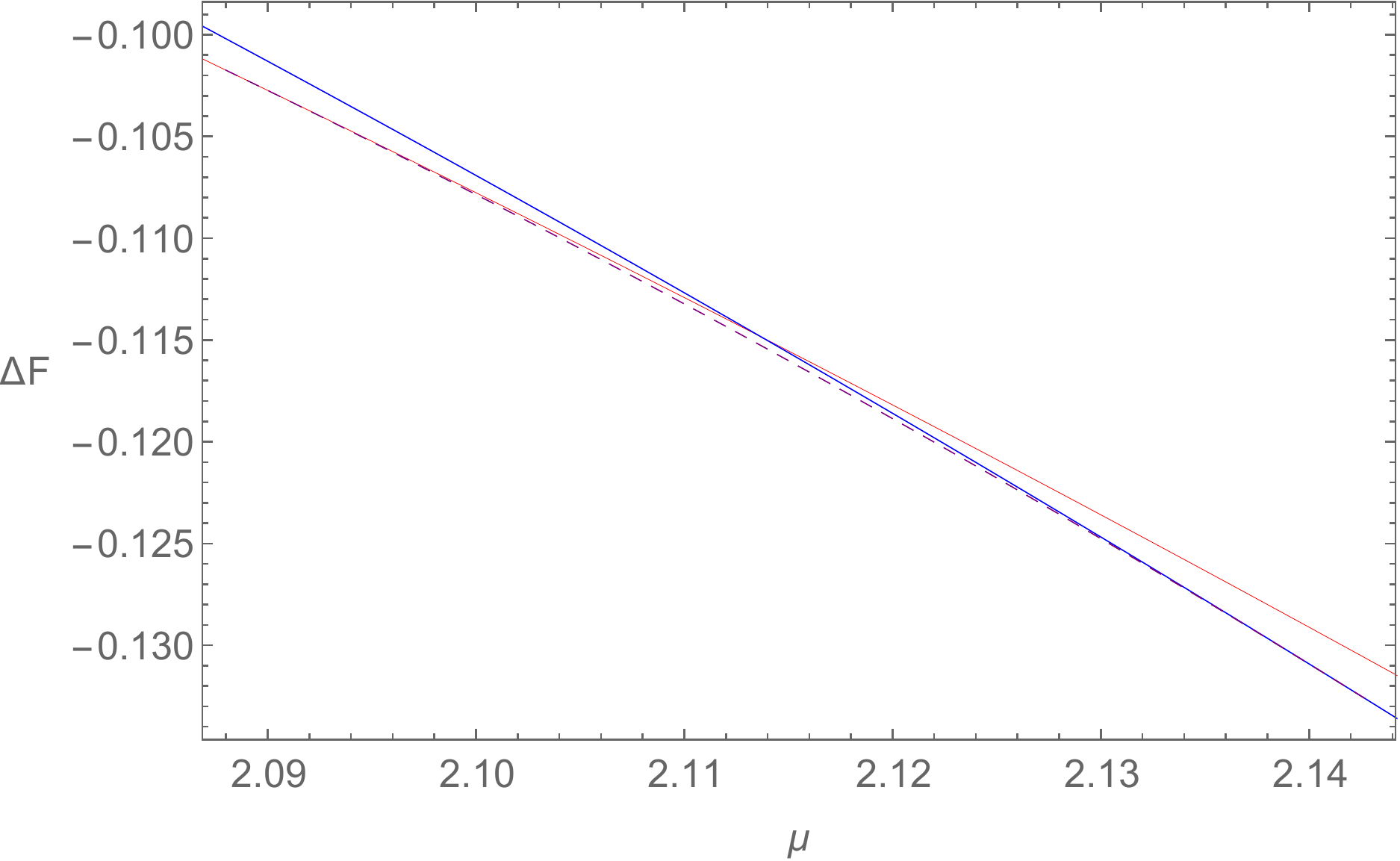}
\end{center}
\caption{The difference of free energy density between the superfluid phases and the vacuum phase. The red line is for S1-Phase, the blue line is for S2-Phase, and the dashed line represents S1+S2-Phase. The right panel is a zoomed-in view in the coexisting phase region, which suggests a phase structure as S1+S2-Phase sandwiched by S1-Phase and S2-Phase.}
\label{energy1}
\end{figure}

\begin{figure}
\begin{center}
\includegraphics[width=7.5cm]{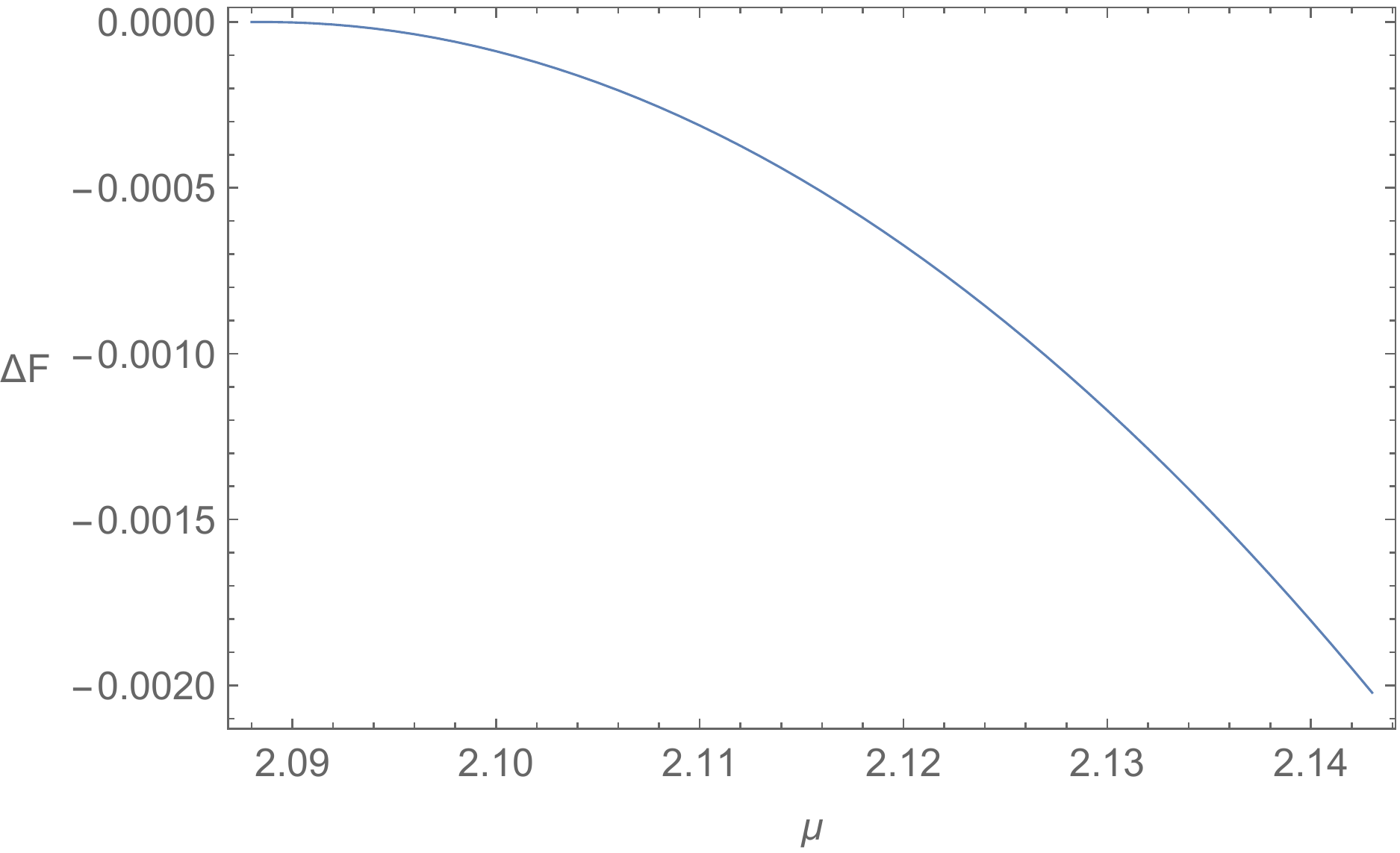}
\includegraphics[width=7.5cm]{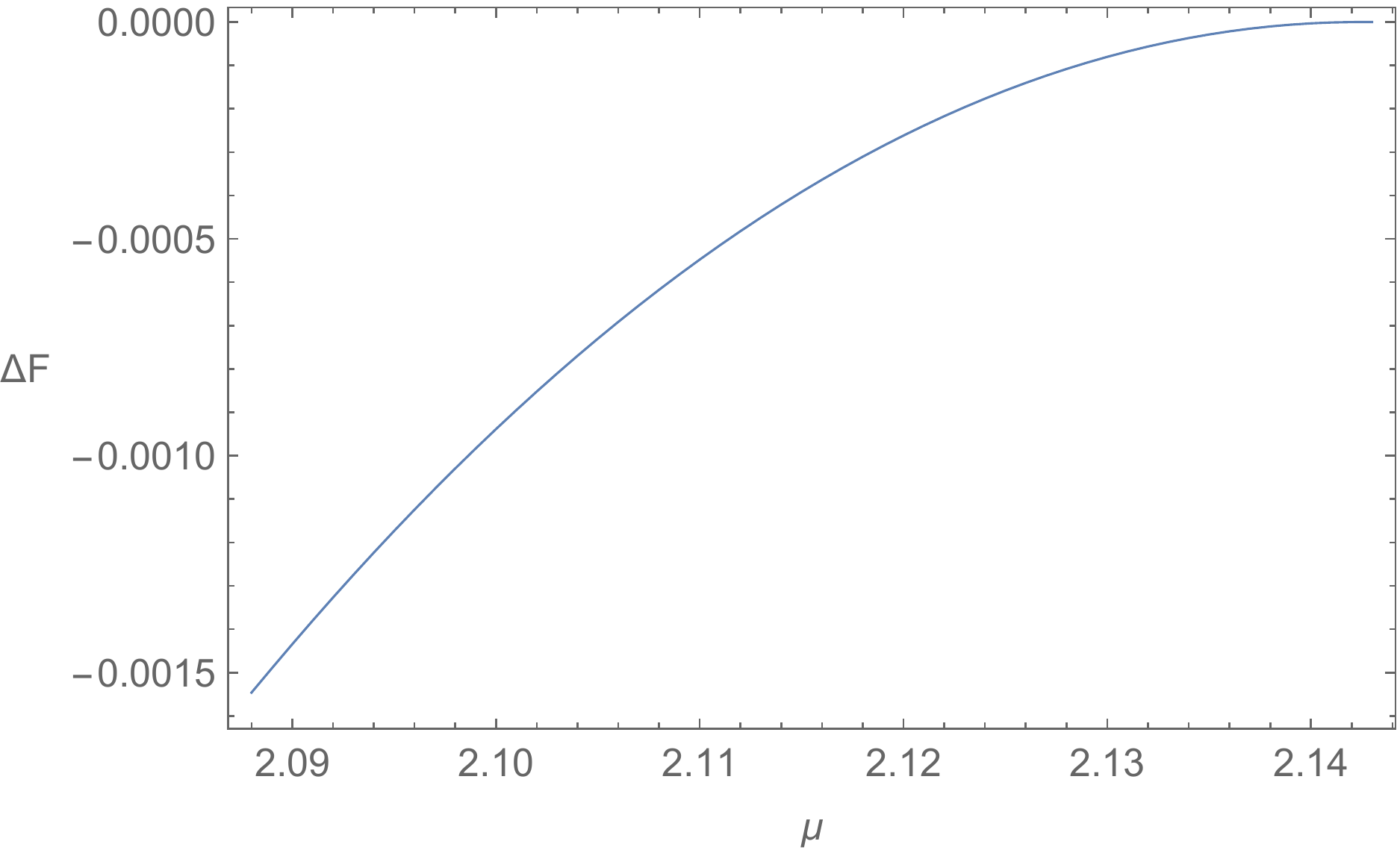}
\end{center}
\caption{The two panels represent the difference of free energy density between S1+S2-Phase and S1-Phase(left)/S2-Phase(right) in the region between $\mu_{c1}$ and $\mu_{c2}$, which indicates that S1+S2-Phase is thermodynamically favored.}
\label{energy2}
\end{figure}

In order to determine whether Figure \ref{phase2} really represents the genuine phase diagram for our two component holographic superfluid model, we are left with one thing to check. Namely we are require to calculate out the corresponding free energy density in the grand canonical ensemble and ensure it is the smallest compared to any other possible configuration. For later notational convenience, we denote the phase only with $\Psi_1$ condensate as S1-Phase, and the phase only with $\Psi_2$ condensate as S2-Phase. While the coexistent phase is denoted as S1+S2-Phase. Now by holography, the free energy density can be obtained from the renormalized on-shell action of bulk matter fields as
\begin{eqnarray}\label{free}
F&=&-\frac{1}{2}[\int dz
\sqrt{-g}i[e\left(\overline{\Psi_1}D_1^b\Psi_1-\Psi_1\overline{D_1^b\Psi_1}\right)+
\left(\overline{\Psi_2}D_2^b\Psi_2-\Psi_2\overline{D_2^b\Psi_2}\right)]A_b\nonumber\\
&&-\sqrt{-h}n_aA_bF^{ab}|_{z=0}]
=-\frac{1}{2}\mu\rho+\int dz\frac{\left(eA_t\psi_1\right)^2}{z^2}+\int dz\left(A_t\psi_2\right)^2,
\end{eqnarray}
where we have taken advantage of the equations of motion as well as the source free boundary conditions for the scalar fields at the AdS boundary.
As revealed in Figure \ref{energy1}, all the superfluid phases give a lower free energy density than that for the vacuum phase. In particular, as shown in Figure \ref{energy2}, S1+S2-Phase has the lowest free energy density compared to S1-Phase and S2-Phase in the coexistent region.

\begin{figure}
\begin{center}
\includegraphics[width=7.5cm]{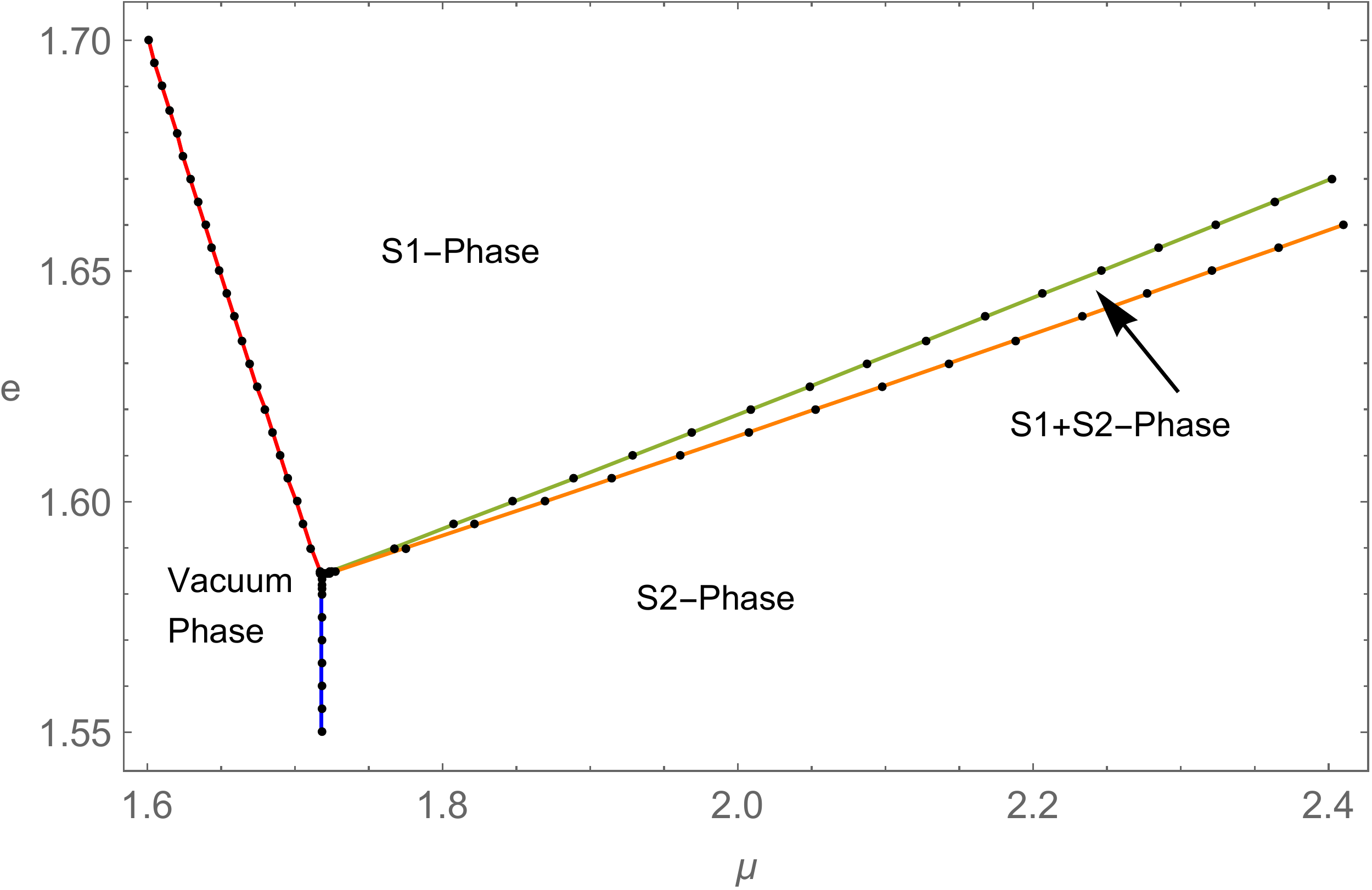}
\end{center}
\caption{The $e-\mu$ phase diagram, where the intersection point is located at $e=1.584$, and $\mu=1.718$.}
\label{pdiagram}
\end{figure}

Using the similar procedure, we can figure out the complete phase diagram in the $e-\mu$ plane by numerics. But before that, we would like to make a wise guess at the rough picture for this phase diagram. First, for the one component holographic superfluid, we have a fixed phase transition point $\mu_2=1.718$ for the phase transition from the vacuum to S2-Phase. Namely the phase boundary between the vacuum and S2-Phase is given by $\mu=\mu_2$ line in the $e-\mu$ plane. While it follows from the scaling symmetry of our holographic system that the phase boundary between the vacuum and S1-Phase is given by the line $\mu=\frac{1.63\times 1.669}{e}=\frac{2.720}{e}$ in the $e-\mu$ plane. So there exists a critical $e_c=\frac{2.720}{u_2}=\frac{2.720}{1.718}=1.584$. When $e<e_c$, $\Psi_2$ starts to condense at $\mu_2$, before $\Psi_1$. In this case, we only have S2-Phase. On the other hand, when $e>e_c$, $\Psi_1$ starts to condense at $\mu_1$, before $\Psi_2$. In this case, as demonstrated for the example $e=1.63$, we shall have S1+S2-Phase sandwiched by S1-Phase and S2-Phase. Actually as plotted in Figure \ref{pdiagram} by numerics, the complete phase diagram is well captured by the above rough guess. Figure \ref{pdiagram} further shows that with a larger $e$ in our holographic model, the coexisting phase will occur in a wider region of the chemical potential, starting from a larger chemical potential.

\section{Optical Conductivity}\label{linearresponse}

In this section, we shall calculate the optical conductivity for our two component holographic superfluid model by the linear response theory acting onto the previous static background solutions. For our purpose, we first rewrite our scalar fields $\psi_1$ and $\psi_2$ in terms of the real and imaginary parts as
\begin{equation}
\psi_1=\psi_{1,r}+i\psi_{1,i},\quad \psi_2=\psi_{2,r}+i\psi_{2,i}.
\end{equation}
Note that the background solution is static and homogeneous, so the bulk perturbation fields $\delta\Phi$ with the following form
\begin{equation}
\delta\Phi\rightarrow \delta\Phi(z)e^{-i\omega t+iqx}
\end{equation}
are decoupled from those modes with a different $\omega$ or a different $q$. Furthermore, the perturbation equations can be reduced to
\begin{eqnarray}
0&=&z(z^3-1)\delta\psi_{1,r}^{''}+(z^3+2)\delta\psi_{1,r}^{'}+z(q^2-\omega^2-e^2A_t^2)\delta\psi_{1,r}\nonumber\\
&&-2ezA_t(e\psi_{1,r}\delta A_t+i\omega\delta\psi_{1,i}),\\
0&=&z(z^3-1)\delta\psi_{1,i}^{''}+(z^3+2)\delta\psi_{1,i}^{'}+z(q^2-\omega^2-e^2A_t^2)\delta\psi_{1,i}\nonumber\\
&&+iez(2\omega A_t\delta\psi_{1,r}+\psi_{1,r}(\omega\delta A_t+q\delta A_x)),\\
0&=&(z^3-1)\delta\psi_{2,r}^{''}+3z^2\delta\psi_{2,r}^{'}+(q^2+z-\omega^2-A_t^2)\delta\psi_{2,r}\nonumber\\
&&-A_t(2\psi_{2,r}\delta A_t+2i\omega\delta\psi_{2,i}),\\
0&=&(z^3-1)\delta\psi_{2,i}^{''}+3z^2\delta\psi_{2,i}^{'}+(q^2+z-\omega^2-A_t^2)\delta\psi_{2,i}\nonumber\\
&&+i(2\omega A_t\delta\psi_{2,r}+\psi_{2,r}(\omega\delta A_t+q\delta A_x)),\\
0&=&z^2(z^3-1)\delta A_t^{''}+3z^4\delta A_t^{'}+(2e\psi_{1,r}^2+2z^2\psi_{2,r}^2+z^2q^2)\delta A_t\nonumber\\
&&+2\psi_{1,r}(2eA_t\delta\psi_{1,r}+i\omega\delta\psi_{1,i})
+2\psi_{2,r}(2A_t\delta\psi_{2,r}+i\omega\delta\psi_{2,i})
+q\omega\delta A_x,\\
0&=&z^2(z^3-1)\delta A_x^{''}+3z^4\delta A_x^{'}+(2e\psi_{1,r}^2+2z^2\psi_{2,r}^2-z^2\omega^2)\delta A_x\nonumber\\
&&-2iq\psi_{1,r}\delta\psi_{1,i}-2iz^2q\psi_{2,r}\delta\psi_{2,i}-z^2q\omega\delta A_t,\label{opticalcond}\\
0&=&2(\psi_{1,r}^{'}\delta\psi_{1,i}-\psi_{1,r}\delta\psi_{1,i}^{'})+
2z^2(\psi_{2,r}^{'}\delta\psi_{2,i}-\psi_{2,r}\delta\psi_{2,i}^{'})+iz^2(\omega\delta A_t^{'}+q\delta A_x^{'}).\label{constraint}
\end{eqnarray}
where we have made use of the fact $\psi_{1,i}=\psi_{2,i}=0$ for the background solution.

To proceed, it is noteworthy that the perturbation of the following form
\begin{equation}
\delta A_t=-\lambda\omega, \delta A_x=\lambda q, \delta\psi_1=e\lambda\psi_1, \delta\psi_2=\lambda\psi_2.
\end{equation}
is essentially kind of gauge transformation
\begin{equation}
A\rightarrow A+\nabla\theta, \psi_1\rightarrow\psi_1e^{ie\theta}, \psi_2\rightarrow\psi_2e^{i\theta}
\end{equation}
with
\begin{equation}
\theta=\frac{1}{i}\lambda e^{-i\omega t+iqx}
\end{equation}
on top of the background solution. This spurious solution generated by the parameter $\lambda$ can be eliminated by gauge fixing. Below we shall choose a gauge such that $\delta A_t=0$ at the AdS boundary. In addition, as we are working with the standard quantization, the Dirichlet boundary conditions will be implemented for $\delta\psi_1$ and $\delta\psi_2$ at the AdS boundary. On the other hand, note that the perturbation equation (\ref{constraint}) turns to be automatically satisfied once the other equations are satisfied, thus we shall forget about (\ref{constraint}) hereafter.

\begin{figure}
\begin{center}
\includegraphics[width=7.5cm]{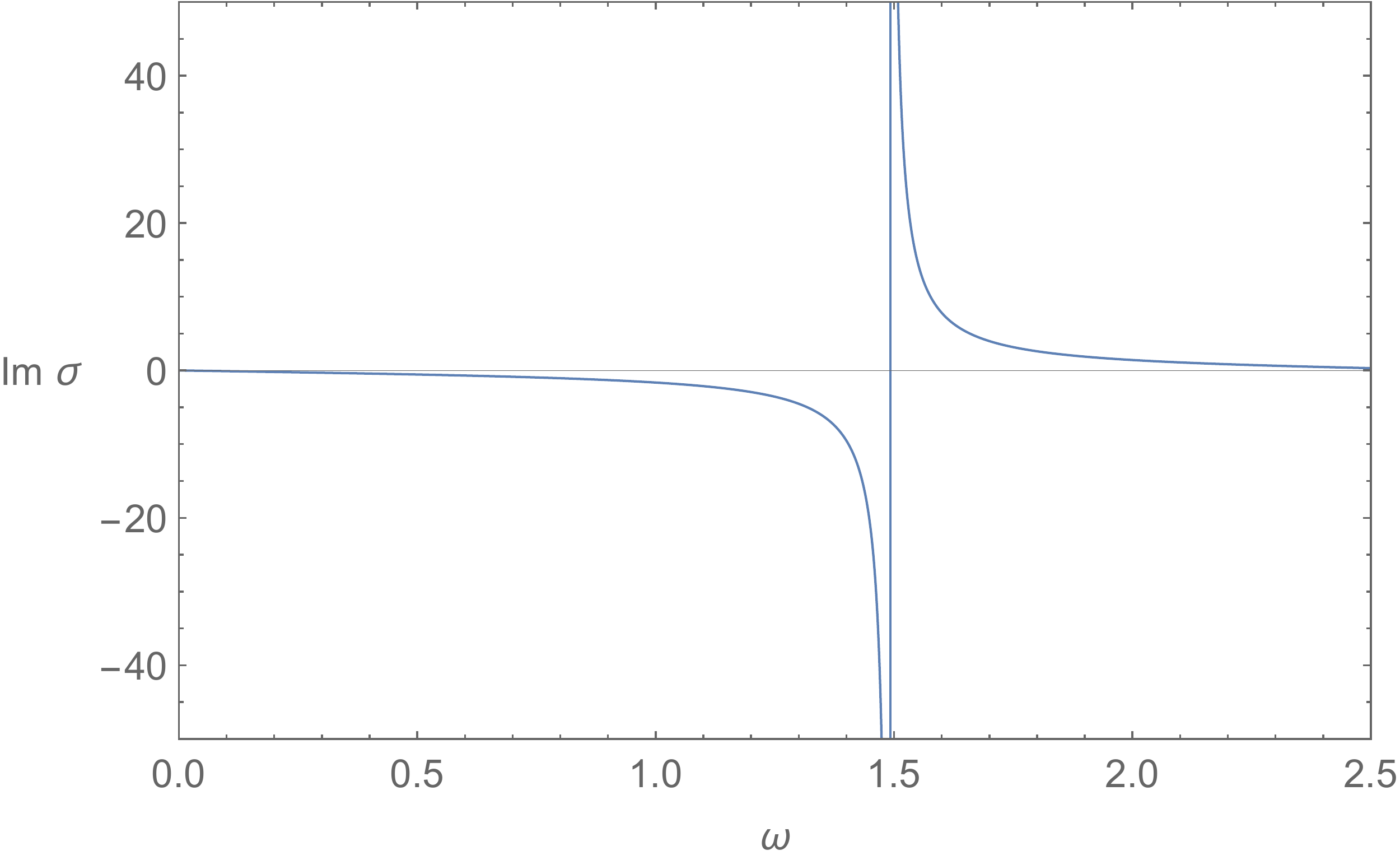}
\includegraphics[width=7.5cm]{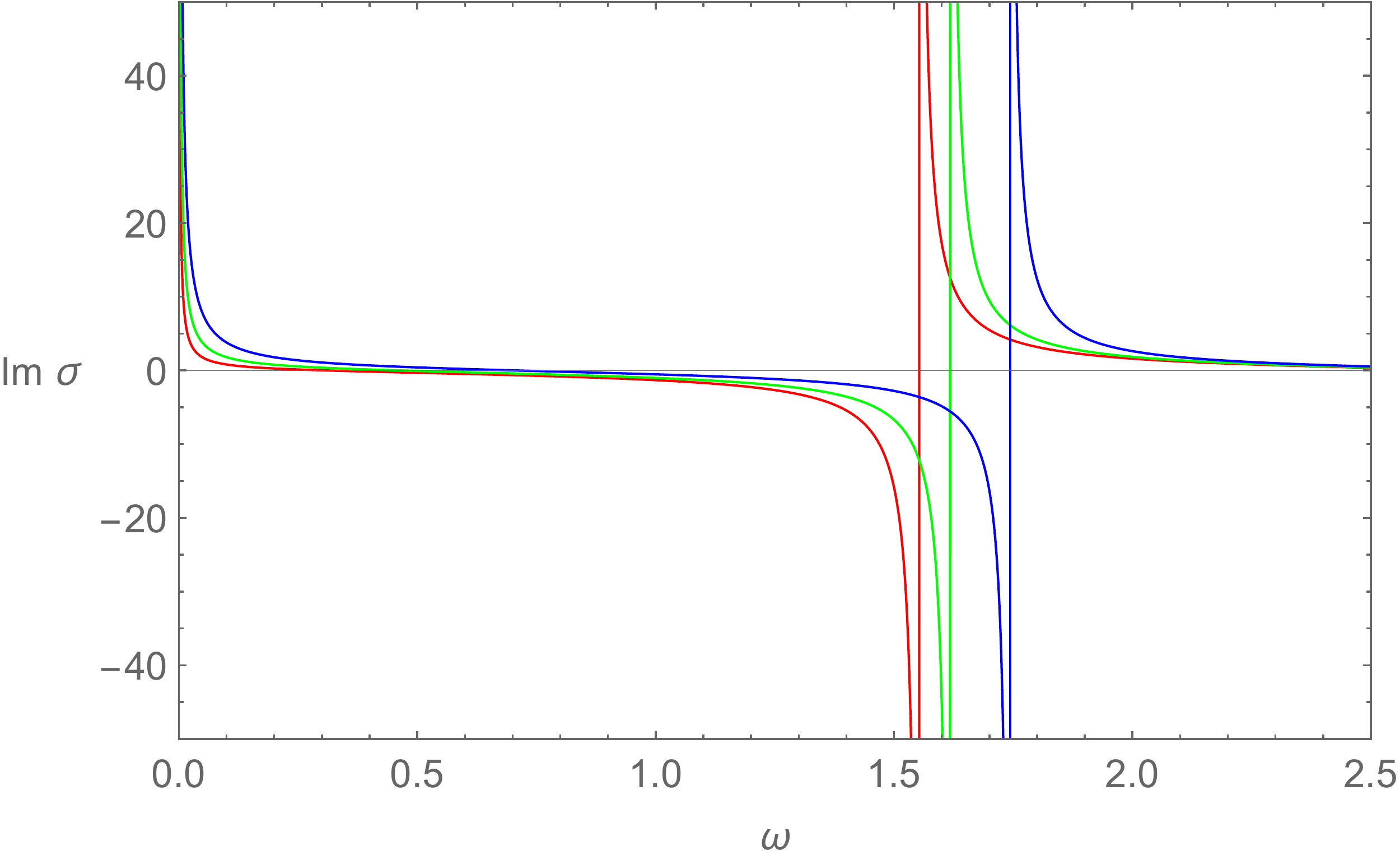}
\end{center}
\caption{The imaginary part of optical conductivity as a function of frequency for the two component holographic superfluid with $e=1.63$. The left panel is for the vacuum phase at $\mu=1.5$, while the right panel is for the S1-Phase at $\mu=1.9$ (Red), S1+S2-Phase at $\mu=2.1$ (Green), and S2-Phase at $\mu=2.3$ (Blue).}
\label{conductivity}
\end{figure}

 With the above preparation, now let us calculate the optical conductivity for our two component holographic superfluid by focusing on the $q=0$ mode. As a consequence, $\delta A_x$ in (\ref{opticalcond}) is decoupled from the other perturbation fields, and can be solved by the pseudo-spectral method.  With the boundary condition $\delta A_x=1$ at the AdS boundary, the holographic optical conductivity can be expressed as
\begin{equation}\label{oc}
\sigma(\omega)=\frac{\partial_z\delta A_x|_{z=0}}{i\omega}.
\end{equation}
Since the real perturbation equation together with the real boundary condition for $\delta A_x$ implies that the real part of the holographic conductivity must vanish, we only depict the nontrivial imaginary part of the optical conductivity in Figure \ref{conductivity} for the vacuum phase and the three superfluid phases. According to the Krames-Kronig relation
\begin{equation}
\mathbf{Im}[\sigma(\omega)]=\frac{1}{\pi}\mathcal{P}\int_{-\infty}^\infty d\omega'\frac{\mathbf{Re}[\sigma(\omega')]}{\omega-\omega'},
\end{equation}
the DC conductivity is equal to zero for the vacuum phase, but acquires a $\delta(\omega)$ peak for all the spontaneous breaking phases due to the $\frac{1}{\omega}$ behavior of the imaginary part of optical conductivity, which tells us that these spontaneous breaking phases correspond to the superfluid phases indeed. Furthermore, the residue for this zero pole is related to the superfluid density as  $\frac{\rho_s}{\mu}$. Thus Figure \ref{conductivity} further indicates that the superfluid density
 becomes large as the chemical potential is cranked up. This is consistent with Figure \ref{phase2}, in which the particle density is increased with the chemical potential. Because as shown in \cite{GNTZ} and \cite{GLNTZ}, $\rho_s=\rho$ at zero temperature. In addition, the other poles give rise to the gapped normal modes for $\delta A_x$. It follows from Figure \ref{conductivity} that the gap becomes larger with the increase of the chemical potential. In the subsequent section, we shall not care about these gapped normal modes any more.

\section{Sound Speed}\label{nmss}
\begin{figure}
\begin{center}
\includegraphics[width=7.5cm]{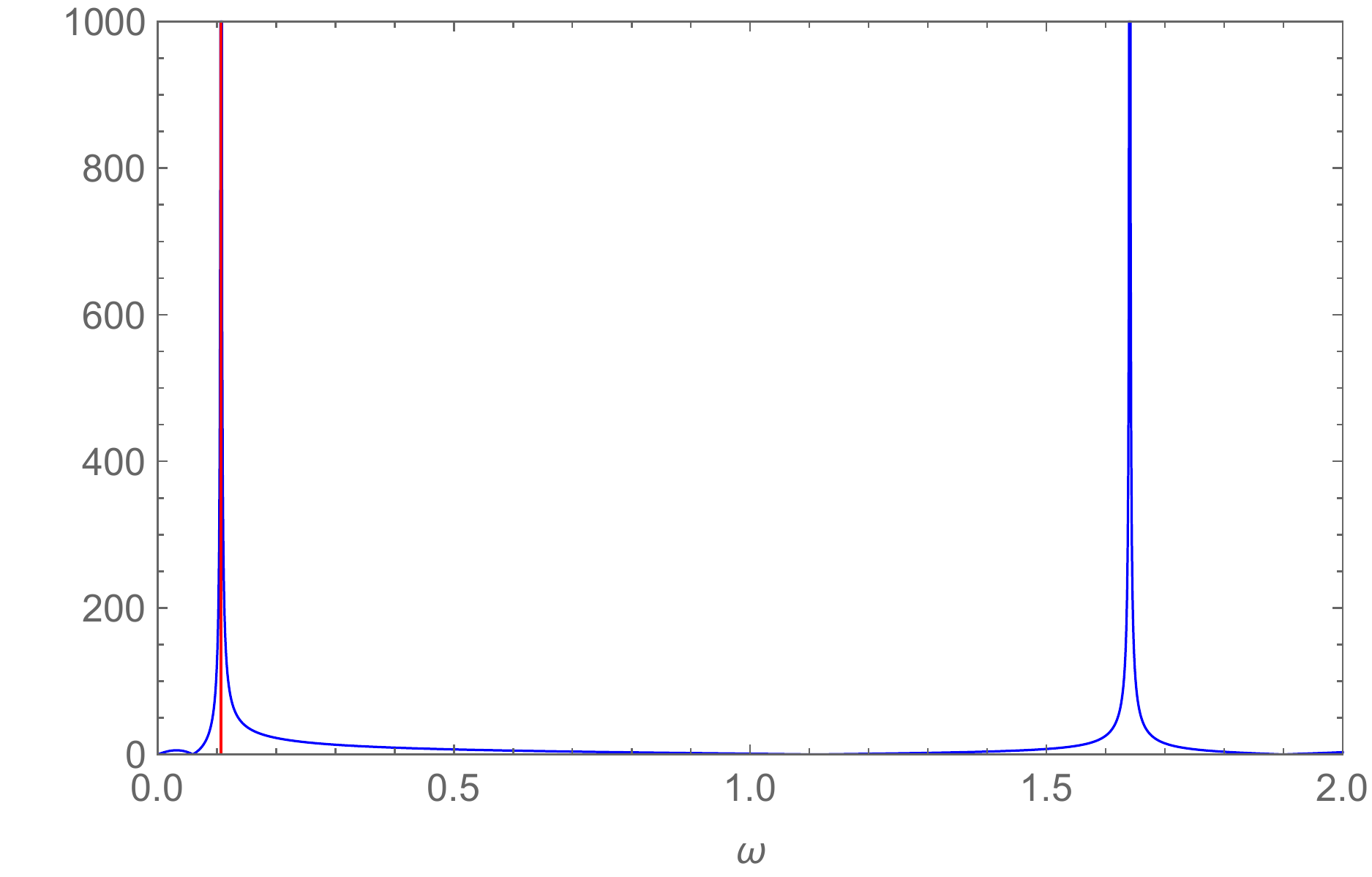}
\end{center}
\caption{The density plot of $|\frac{det[\mathcal{L}(\omega)]'}{det[\mathcal{L}(\omega)]}|$ with $q=0.3$ and $e=1.63$ for S1+S2-Phase at $\mu=2.1$. The normal modes give rise to the peaks, where the red line pins down the hydrodynamic normal mode at $\omega_0\approx0.107$.}
\label{densityplot}
\end{figure}

In order to calculate the speed of sound by the linear response theory for our two component holographic superfluid model, we are required to work on the hydrodynamic normal modes of the gapless Goldstone boson from the spontaneous symmetry breaking in the superfluid phases. To this end, we need work not only with the $q=0$ mode, but also with the $q\neq 0$ modes. In addition, we replace $\delta A_x=1$ by $\delta A_x=0$ at the AdS boundary. Furthermore we massage the linear perturbation equations as well as the boundary conditions into the form $\mathcal{L}(\omega)v= 0$ with $v$ the values of perturbation fields at the grid points associated with the pseudo-spectral method. Note that the normal modes satisfy the condition $det[\mathcal{L}(\omega)]= 0$, so these modes can be spotted in the density plot $|\frac{det[\mathcal{L}(\omega)]'}{det[\mathcal{L}(\omega)]}|$ with the prime the derivative with respect to $\omega$ here. As a demonstration, we depict the corresponding density plot for the S1+S2-Phase at $\mu=2.1$ with $q=0.3$ and $e=1.63$ in Figure \ref{densityplot}, where the hydrodynamic normal mode locates at the closest peak to the origin, marked in red line.

\begin{figure}
\begin{center}
\includegraphics[width=7.5cm]{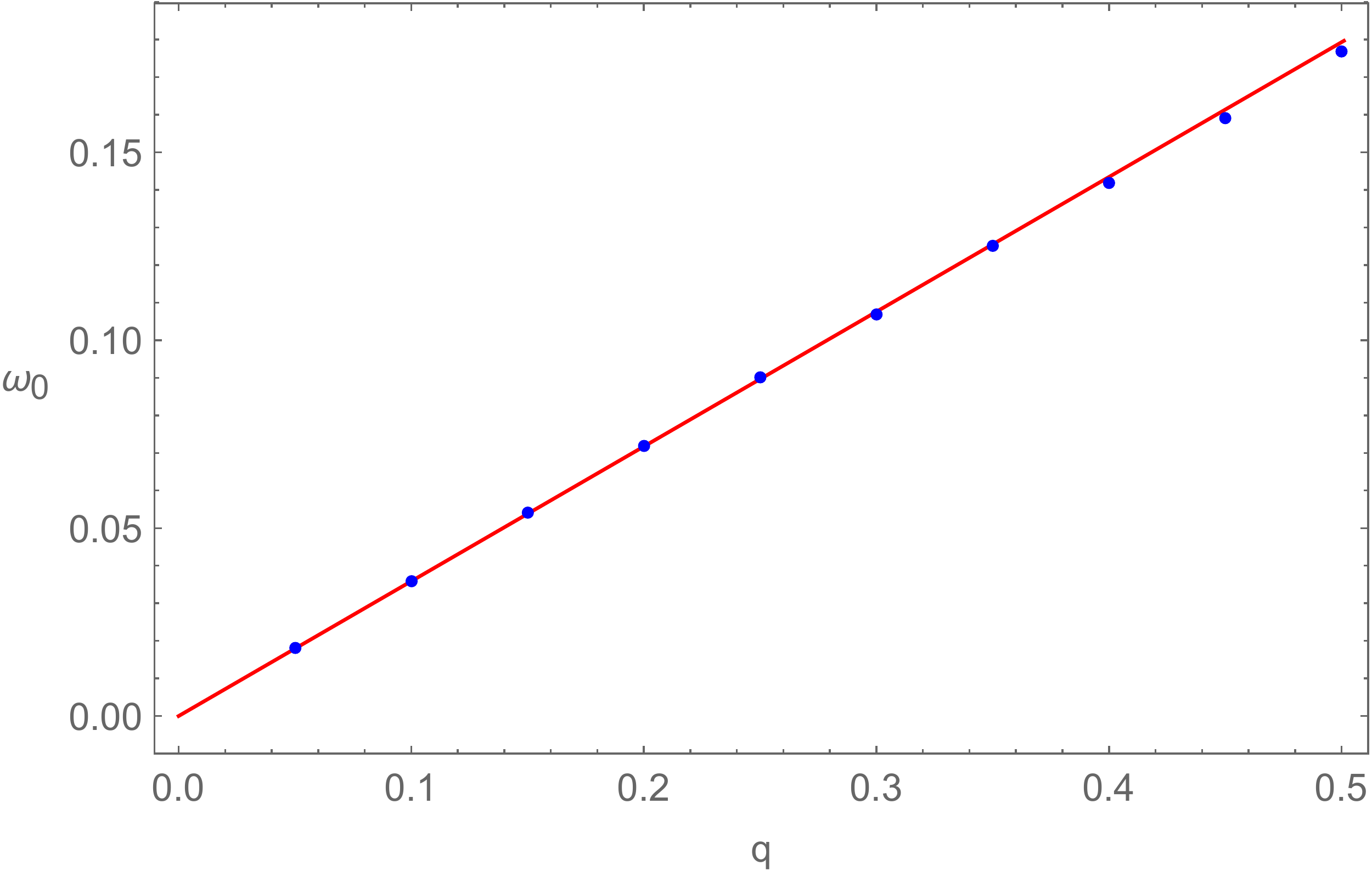}
\end{center}
\caption{ The dispersion relation of the gapless Goldstone mode for S1+S2 phase at $\mu=2.1$. By the fitting formula $\omega_0=v_sq$, the sound speed can be extracted as $v_s=0.359$.}
\label{dispersion}
\end{figure}

\begin{figure}
\begin{center}
\includegraphics[width=7.5cm]{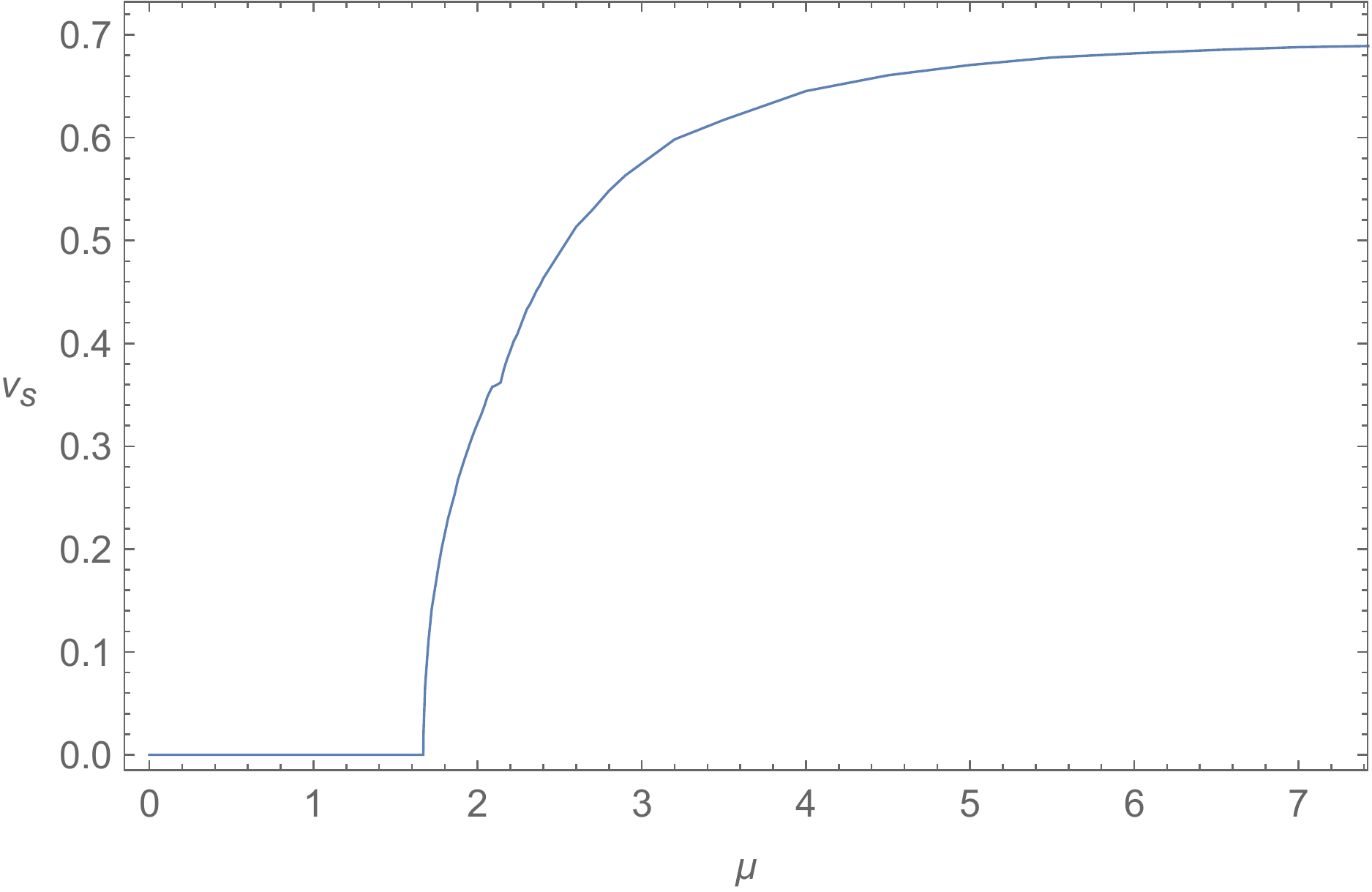}
\end{center}
\caption{ The variation of sound speed with respect to the chemical potential. The phase transitions turn out to be also signaled by the non-smoothness of sound speed at the critical points. In addition, the sound speed approaches $\frac{1}{\sqrt{2}}$ at the large chemical potential limit. }
\label{ss}
\end{figure}

With this strategy, we further plot the dispersion relation for the gapless Goldstone mode in Figure \ref{dispersion} for S1+S2-Phase at $\mu=2.1$. Then we can extract the sound speed $v_s$ by the fitting formula $\omega_0=v_s q$ for small $q$s. We present the variation of sound speed with respect to the chemical potential in Figure \ref{ss}. As one can see, the sound speed increases with the chemical potential. Furthermore, Figure \ref{ss} indicates that the sound speed can also be used to signal the phase transition because the phase transition leaves its footprint by making the sound speed also non-smooth at the critical point. In addition, when the chemical potential is much larger than the confining scale, the sound speed saturates to a constant, which is in good agreement with the predicted value $\frac{1}{\sqrt{2}}$ from conformal field theory for the superfluid condensate with the conformal dimension $2$\cite{HKS,Y,HY}. This is reasonable since the conformality is believed to be restored in large chemical potential limit.

\section{Conclusion}\label{end}
We have initiated the study of two component superfluid at zero temperature by investigating its dual gravitational system with two complex scalar fields coupled to a single U(1) gauge field in the AdS soliton background. We make a qualitative analysis on the possible existence of the two-component superfluid phase and verify its occurrence by numerically constructing the bulk solutions as well as making an analysis of the corresponding free energy density. We further figure out the complete phase diagram in the $e-\mu$ plane with the coexistent region specified. In addition, we make use of the linear response theory to work out the optical conductivity and sound speed. The onset of pole of optical conductivity at $\omega=0$ for the spontaneous breaking phase confirms its superfluid phase interpretation. On the other hand, the resulting sound speed exhibits the non-smoothness at the critical points. Thus we can also use the behavior of sound speed to identify the onset of phase transitions. Furthermore, as expected from the boundary conformal field theory, our sound speed approaches $\frac{1}{\sqrt{2}}$ at the large chemical potential.

Finally, note that we have restricted ourselves onto the the probe limit, so we would like to conclude with some comments on the back reaction effect, which is controlled parametrically by $e_2$.  As suggested in \cite{HW} for one component holographic superfluid model, the phase diagram is expected to have no essential change at zero temperature when $e_2$ is sufficiently large. But when $e_2$ is decreased, the phase transition from the hairless AdS soliton to the hairy AdS soliton may become first order. If one decreases $e_2$ further, there may be a hairy AdS black hole phase emergent between the hairless AdS soliton and the hairy AdS soliton. It is intriguing to work out such a fully backreacted problem explicitly for our two component holographic superfluid model and figure out how the AdS soliton geometry (hairless or hairy) is connected with AdS black hole geometry (hairless or hairy) in the full phase diagram not only at zero temperature but also at finite temperature. The probe limit we have worked out is supposed to provide us with a good preparation to attack this fully backreacted problem. We hope to report the relevant result elsewhere in the near future.

\begin{acknowledgments}
R.L. and J.Z. are supported by NSFC with Grant No.11205048. R.L. is also supported by the Foundation
for Young Key Teacher of Henan Normal University.
Y.T. is partially supported by NSFC with Grant No.11475179 and the Opening Project of Shanghai Key Laboratory of High Temperature Superconductors(14DZ2260700).
H.Z. is supported in part by the Belgian Federal
Science Policy Office through the Interuniversity Attraction Pole
P7/37, by FWO-Vlaanderen through the project
G020714N, and by the Vrije Universiteit Brussel through the
Strategic Research Program ``High-Energy Physics''. He is also an individual FWO Fellow supported by 12G3515N.
This work is also partially supported by ``the Fundamental Research Funds for the Central Universities" with Grant No.2015NT16.

\end{acknowledgments}


\begin{thebibliography}{20}


\bibitem{M}J. M. Maldacena, The Large-N Limit of Superconformal Field Theories and Supergravity, Adv. Theor. Math. Phys, 2, 231(1998)[Int. J. Theor. Phys. 38, 1113(1999)].
\bibitem{GKP}S. S. Gubser, I. R. Klebanov, and A. M. Polyakov, Gauge Theory Correlators from Non-Critical String Theory, Phys. Lett. B 428, 105(1998).
\bibitem{W}E. Witten, Anti-de Sitter Space and Holography, Adv. Theor. Math. Phys. 2, 253(1998).
\bibitem{G}S. S. Gubser, Breaking an Abelian Gauge Symmetry near a Black Hole Horizon, Phys. Rev. D 78, 065034(2008).
\bibitem{HHH1}S. A. Hartnoll, C. P. Herzog, and G. T. Horowitz, Building a Holographic Superconductor, Phys. Rev. Lett. 101, 031601(2008).
\bibitem{HHH2}S. A. Hartnoll, C. P. Herzog, and G. T. Horowitz, Holographic Superconductors, JHEP 0812, 015(2008).
\bibitem{GP}S. S. Gubser and S. S. Pufu, The Gravity Dual of a P-Wave Superconductor, JHEP 0811, 033(2008).
\bibitem{DG}A. Donos and J. P. Gauntlett, Holographic Helical Superconductors, JHEP 1112, 091(2011).
\bibitem{CLL}R. G. Cai, L. Li, and L. F. Li, A Holographic P-Wave Superconductor Model, JHEP 1401, 032(2014).
\bibitem{CKMWY}J. W. Chen, Y. J. Kao, D. Maity, W. Y. Wen, and C. P. Yeh, Towards A Holographic Model of D-Wave Superconductors, Phys. Rev. D 81, 106008(2010).
\bibitem{BHRY}F. Benini, C. P. Herzog, R. Rahman, and A. Yarom, Gauge Gravity Duality for D-Wave Superconductors: Prospects and Challenges, JHEP 1011, 137(2010).
\bibitem{NRT}T. Nishioka, S. Ryu, and T. Takayanagi, Holographic Superconductor/Insulator Transition at Zero Temperature, JHEP 1003, 131(2010).
\bibitem{ZHTC}X. Zhang, C. L. Hung, S. K. Tung, and C. Chin, Science 335, 1070(2012).
\bibitem{HW}G. T. Horowitz and B. Way, Complete Phase Diagrams for a Holographic Superconductor/Insulator System, JHEP 1011, 011(2010).
\bibitem{CLZ}R. G. Cai, H. F. Li, and H. Q. Zhang, Analytical Studies on Holographic Insulator/Superconductor Phase Transitions, Phys. Rev. D 83, 126007(2011).
\bibitem{BHMRS}P. Basu, J. He, A. Mukherjee, M. Rozali, and H. H. Shieh, Competing Holographic Orders, JHEP 1010, 092(2010).
\bibitem{CLLW}R. G. Cai, L. Li, L. F. Li, and Y. Q. Wang, Competition and Coexistence of Order Parameters in Holographic Multi-Band Superconductors, JHEP 1309, 074(2013).
\bibitem{HLM}C. Y. Huang, F. L. Lin, and D. Maity, Holographic Multi-Band Superconductor, Phys. Lett. B 703, 633(2011).
\bibitem{DGSW}A. Donos, J. P. Gauntlett, J. Sonner, and B. Withers, Competing Orders in M-theory: Superfluids, Stripes and Metamagnetism, JHEP 1303, 108(2013).
\bibitem{DM}D. Musso, Competition/Enhancement of Two Probe Order Parameters in the Unbalanced Holographic Superconductor, JHEP 1306, 083(2013).
\bibitem{KKS}A. Krikun, V. P. Kirilin, and A. V. Sadofyev, Holographic Model of the $S^{\pm}$ Multiband Superconductor, JHEP 1307, 136(2013).
\bibitem{NCGZ}Z. Y. Nie, R. G. Cai, X. Gao, and H. Zeng, Competition between the S-Wave and P-Wave Superconductivity Phases in a Holographic Model, JHEP 1311, 087(2013).
\bibitem{LCLW}L. F. Li, R. G. Cai, L. Li, and Y. Q. Wang, Competition between S-Wave Order and D-Wave Order in Holographic Superconductors, JHEP 1408, 164(2014).
\bibitem{N}M. Nishida, Phase Diagram of a Holographic Superconductor Model with S-Wave and D-Wave, JHEP 1409, 154(2014).
\bibitem{CB}P. Chaturvedi and P. Basu, Holographic Quantum Phase Transitions and Interacting Bulk Scalars, Phys. Lett. B 739, 162(2014).
\bibitem{AAJML}I. Amado, D. Arean, A. Jimenez-Alba, L. Melgar, and I. S. Landea, Holographic S+P Superconductors, Phys. Rev. D 89, 026009(2014).
\bibitem{WYW}W. Y. Wen, M. S. Wu, and S. Y. Wu, A Holographic Model of Two-Band Superconductor, Phys. Rev. D 89, 066005(2014).
\bibitem{MRM}D. Momeni, M. Raza, and R. Myrzakulov, Analytical Coexistence of S, P, S+P Phases of a Holographic Superconductor, Int. J. Geom. Methods Mod. Phys. 12, 1550048(2015).
\bibitem{NCGLZ}Z. Y. Nie, R. G. Cai, X. Gao, L. Li, and H. Zeng, Phase Transitions in a Holographic S+P Model with Backreaction, Eur. Phys. J. C 75, 559(2015).
\bibitem{NZ}Z. Y. Nie and H. Zeng, P-T Phase Diagram of a Holographic S+P Model from Gauss-Bonnet Gravity, JHEP 1510, 047(2015).
\bibitem{MSW}M. S. Wu, S. Y. Wu, and H. Q. Zhang, Vortex in Holographic Two-Band Superfluid/Superconductor, arXiv:1511.01325[hep-th].
\bibitem{CLLY}R. G. Cai, L. Li, L. F. Li, and R. Q. Yang, Introduction to Holographic Superconductor Models, Sci. China Phys. Mech. Astron 58, 060401(2015).
\bibitem{Helium1}J. Tuoriniemi, J. Martikainen, E. Pentti, A. Sebedash, S. Boldarev, and G. Pickett,  Towards Superfluidity of
$^3$he Diluted by $^4$he, Journal of Low Temperature Physics 129, 531(2002).
\bibitem{Helium2}J. Rysti, J. Tuoriniemi, and A. Salmela,  Effective $^3$he
Interactions in Dilute $^3$he-$^4$he Mixtures, Phys. Rev. B 85, 134529(2012).
\bibitem{Cold1}F. Schreck, L. Khaykovich, K. L. Corwin, G. Ferrari,
T. Bourdel, J. Cubizolles, and C. Salomon, Quasipure
Bose-Einstein Condensate Immersed in a Fermi Sea, Phys.
Rev. Lett. 87, 080403(2001).
\bibitem{Cold2}I. Ferrier-Barbut, M. Delehaye, S. Laurent, A. T. Grier, M. Pierce, B. S. Rem, F. Chevy, and C. Salomon, A Mixture of Bose and Fermi Superfluids, Science 345, 6200(2014).
\bibitem{footnote}Hereafter we shall restrict ourselves onto the holographic interpretation of superfluids.
\bibitem{JE}J. Erdmenger, X. H. Ge, and D. W. Pang, Striped Phases in the Holographic Insulator/Superconductor Transition, JHEP 1311, 027(2013).
\bibitem{KWG}X. M. Kuang, B. Wang, and X. H. Ge, Observing the Inhomogeneity in the Holographic Models of Superconductors, Mod. Phys. Lett. A 29, 1450070(2014).


\bibitem{GNTZ}M. Guo, C. Niu, Y. Tian, and H. Zhang, Applied AdS/CFT with Numerics, PoS Modave 2015, 003(2016).
\bibitem{GLNTZ}M. Guo, S. Lan, C. Niu, Y. Tian, and H. Zhang, Note on Zero Temperature Holographic Superfluids, Class. Quant. Grav. 33, 127001(2016).


\bibitem{HKS}C. P. Herzog, P. K. Kovtun, and D. T. Son, Holographic Model of Superfluidity, Phys. Rev. D 79, 066002(2009).
\bibitem{Y}A. Yarom, Fourth Sound of Holographic Superfluids, JHEP 0907, 070(2009).
\bibitem{HY}C. P. Herzog and A. Yarom, Sound Modes in Holographic Superfluids, Phys. Rev. D 80, 106002(2009).

\end{thebibliography}
\end{document}